\documentstyle[prl,aps,psfig,floats]{revtex}

\begin{document}  

\twocolumn[\hsize\textwidth\columnwidth\hsize\csname 
@twocolumnfalse\endcsname 

\title{\vspace*{-1cm}\hfill 
{\tt Phys.~Rev.~Lett., in print}
       \vspace{0.5cm}\\ 
Clean and As-covered zinc-blende GaN (001) surfaces: 
Novel surface structures and surfactant behavior}

\author{J\"org Neugebauer, Tosja Zywietz, Matthias Scheffler}
\address{Fritz-Haber-Institut der Max-Planck-Gesellschaft, Faradayweg
4--6, D-14195 Berlin, Germany}
\author{John E. Northrup and Chris G. Van de Walle}
\address{
Xerox Palo Alto Research Center, 3333 Coyote Hill Road, Palo Alto, California 94304}
\date{\today}

\maketitle

\begin{abstract}
  We have investigated clean and As-covered zinc-blende GaN (001)
  surfaces, employing first-principles total-energy calculations.  For
  clean GaN surfaces our results reveal a novel surface structure very
  different from the well-established dimer structures commonly observed
  on polar III-V (001) surfaces: The energetically most stable surface
  is achieved by a Peierls distortion of the truncated $(1\times1)$
  surface rather than through addition or removal of atoms. This
  surface exhibits a $(1\times4)$ reconstruction consisting of {\em
  linear Ga tetramers}.  Furthermore, we find that a submonolayer of
  arsenic significantly lowers the surface energy indicating that As
  may be a good surfactant.  Analyzing surface energies and band
  structures we identify the mechanisms which govern these unusual
  structures and discuss how they might affect growth properties.
\end{abstract}

]

Its wide direct bandgap and strong chemical bonds render GaN an
ideal material for optoelectronic devices in the blue/UV region of the
optical spectrum. Recently, the fabrication of highly efficient blue
LEDs\cite{nakamura94} and prototypes of a blue laser have been
reported~\cite{nakamura97}. However, despite progress in device
fabrication an understanding of the fundamental growth mechanisms is
still in its infancy, and even the atomic structure of the surface is
not well understood. Only recently atomically resolved scanning
tunneling micrographs have been obtained for wurtzite GaN
surfaces~\cite{asmith97}. A detailed knowledge of the properties and
structure of these surfaces is crucial to improve growth in a
systematic way.

The stable crystal phase of GaN is the wurtzite structure.  However,
cubic (zinc-blende) GaN can be grown epitaxially on cubic SiC or GaAs.
Cubic GaN exhibits a number of properties very appealing for device
applications: It has a lower bandgap than the wurtzite phase (by
0.2\,eV) and can be easily cleaved.  Cubic GaN has been grown e.g. on
cubic GaAs\,(001) substrates \cite{brandt95,brandt96,schikora96}.
Growing on this substrate Brandt {\it et al.}  observed in going from
N-rich conditions to Ga-rich conditions a reversible sequence of
reconstructions exhibiting $(1\times1)$, $(2\times2)$, and
$c(2\times2)$ reflection high energy electron diffraction (RHEED)
patterns~\cite{brandt95}. The $c(2\times2)$ and $(2\times2)$
reconstructions have been also reported by other
groups~\cite{schikora96}. STM measurements further revealed that the
$(2\times2)$ structure contains one dimer per surface cell; but the
chemical nature of the dimer could not be
resolved~\cite{wassermeier97}.  However, recently Feuillet {\it et
al.}\cite{feuillet97} observed a $(1\times4)$ (N rich) and a
$(1\times1)$ (Ga rich) reconstruction for GaN(001) grown on cubic SiC.
Only when exposing these surfaces to an As background pressure the two
surface reconstructions commonly found for GaN on GaAs [$(2\times2)$
and $c(2\times2)$] were observed. Based on these results it appears
that the GaN\,(001) $(2\times2)$ and $c(2\times2)$ structures obtained
in growth on GaAs substrates are stabilized by As adsorption or
segregation, but that the $(1\times4)$ is an intrinsic reconstruction
of the clean surface~\cite{feuillet97}.

In this Letter we address these open questions, which are relevant for
understanding the surface properties and growth mechanisms, by
detailed density functional theory calculations. We have performed
calculations for an extensive set of possible structures for clean and
As-covered GaN\,(001) surfaces. On the basis of these results we
conclude that the $(1\times4)$ structure is an intrinsic
reconstruction of GaN\,(001) consisting of linear Ga-tetramers. The
$(2\times2)$ structure observed for GaN grown on GaAs\,(001) cannot be
understood as an intrinsic reconstruction. Instead, we propose that
1/2 monolayer of As-As dimers gives rise to this phase.

Let us first start with a few general remarks concerning cubic (001)
semiconductor surfaces. At the ideal abrupt III-V (001) surface each
surface atom has two dangling bonds.  For cations (anions) each
dangling bond is occupied with 3/4 (5/4) electrons. The high density
of partially occupied surface dangling bonds renders such a surface
very unstable and surface reconstruction occurs.  Surface
reconstructions on semiconductor surfaces are commonly driven by:
$(i)$ reducing the number of dangling bonds [on conventional (001)
surfaces this is realized by dimer and missing dimer formation],
$(ii)$ minimizing the electronic energy (this is commonly formulated
as the electron counting rule: All energetically low-lying anion
dangling-bond states are doubly occupied, all cation dangling-bond
states are empty), and $(iii)$ minimizing the electrostatic energy by
optimizing the arrangement of the charged surface atoms. Despite their
simplicity these rules have been very succesful in explaining the
structure of polar and non-polar surfaces for a wide variety of
semiconductors.

In accordance with rule $(i)$ common (001) surface reconstructions
involve a combination of dimers and missing dimers. The only exception
of this rule has been reported for the InP\,(001) surface where
trimer units have been observed by STM.\cite{macpherson96} The
stability of dimers can be understood noting the specific topology of
(001) surfaces. Each surface atom is two-fold coordinated and has two
backbonds. Thus, dimers can be formed simply by {\em rotating} the
surface atoms. {\em Stretching} or {\em breaking} of back bonds (which
is energetically unfavorable) is not required. In this letter we will
show that the energetically most favorable structure for GaN\,(001)
does {\em not} consist of dimers.  A new and hitherto never reported
structure consisting of linear Ga-tetramers is found to be most
stable. We explain this unexpected behavior of GaN surfaces in terms
of the large atomic mismatch between Ga and N and the very different
cohesive and binding energies of the two species.

We have performed calculations of the total energy and atomic
structure employing the local-density approximation (LDA) and the
first-principles pseudopotential approach. The relative stability of
possible surface structures has been determined within the
thermodynamically allowed range of the Ga-chemical potential:
$\mu_{\rm Ga(bulk)}-\Delta H_{\rm GaN}<\mu_{\rm Ga}<\mu_{\rm
Ga(bulk)}$. $\Delta H_{\rm GaN}$ is the formation enthalpy of bulk GaN
with respect to Ga-bulk and N$_2$ molecules. The calculations have
been performed with the Ga $3d$ electrons treated explicitly as
valence electrons and with a plane wave cutoff of 60\,Ry. Convergence
checks showed that treating the Ga $3d$ electrons as valence states is
crucial for calculating surface energies. Describing the 3$d$
electrons in the non-linear core correction (nlcc) which would be
computationally much less demanding is not sufficient: A detailed
analysis showed that the nlcc approximation for Ga systematically
overestimates the strength of Ga-Ga bonding relative to Ga-N
bonding. This explains, e.g., why the nlcc approximation severely
underestimates the formation enthalpy $\Delta H_{\rm GaN}$ of GaN
[0.5\,eV compared to 0.95\,eV ($3d$ included) and 1.1\,eV
experiment]. For details of the method we refer to
Refs.~\onlinecite{prb1_94,northrup96,cpc97,mrs95_tb_s}.

The surfaces were modeled by repeated slabs with two equivalent
surfaces on each side. The slabs consisted of 9 layers of
GaN. Tests performed with slabs containing 15 layers showed that the
9-layer slabs are adequate.  Relaxation of the atoms in the top two
layers of each side of the slab was found to be sufficient. A
$(4\times4)$ Monkhorst-Pack mesh was used to sample the Brillouin
zone~\cite{monkhorst76}.

\setlength{\unitlength}{1mm}
\begin{figure}[tb]
 \begin{picture}(90,45)(0,-3)
    \psfig{file=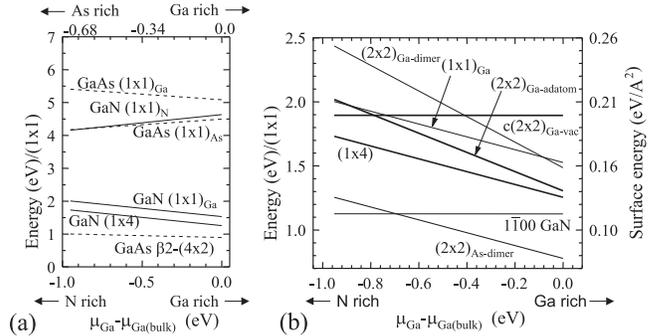,width=8.5cm,clip=f}
 \end{picture}
\caption{ Surface energies for (a) $(1\times1)$ GaN (solid lines) and
GaAs (dashed lines) surfaces and (b) for clean and As covered
reconstructed GaN surfaces as a function of the Ga chemical potential
$\mu_{\rm Ga}$. Only the thermodynamically allowed range is shown. The
surface structures are described in the text. In (a) also the
structures with the lowest surface energy have been included: for GaAs
the $\beta2-(4\times2)$ and for GaN the $(1\times4)$ tetramer
structure [see Fig. 3(a)]. Also included is the surface energy of the
non-polar GaN $(1{\bar 1}00)$ from Ref.~[10]. Note that the
thermodynamically allowed range in (a) is different for GaAs (top
axis) and GaN (bottom axis).}
\label{fig:form_erg}
\end{figure}

We will at first focus on the so-called ``ideal'' surface geometry,
the unreconstructed $(1\times1)$ structure. The calculated surface
energies per $(1\times1)$ cell are shown in
Fig.~\ref{fig:form_erg}. Also included are the calculated surface
energies for GaAs, as reference to a ``traditional'' semiconductor.
Note that for GaAs {\em both} the Ga and As terminated $(1\times1)$
surfaces are energetically highly unfavorable with the Ga- terminated
surface slightly higher in energy than the As-terminated
surface. Compared to the energetically favored $\beta2(4\times2)$ GaAs
surface\cite{northrup93b} the unreconstructed structures are about 5
times higher in energy [see Fig.~1(a)]. For GaN we find a
strikingly different behavior: While the N-terminated surface is close
in energy to the As-terminated GaAs surface the Ga-terminated surface
is more than\,3.5 eV per $(1\times1)$ cell lower in energy than the
corresponding GaAs surface.  In fact, as will be shown below the
energetically most stable GaN $(1\times4)$ surface is only slightly
more stable [by $\approx0.27\,$eV per $(1\times1)$ cell] than the
``ideal'' Ga-terminated GaN surface.

\setlength{\unitlength}{1mm}
\begin{figure}[tb]
 \begin{picture}(70,50)(0,-3)
   \psfig{file=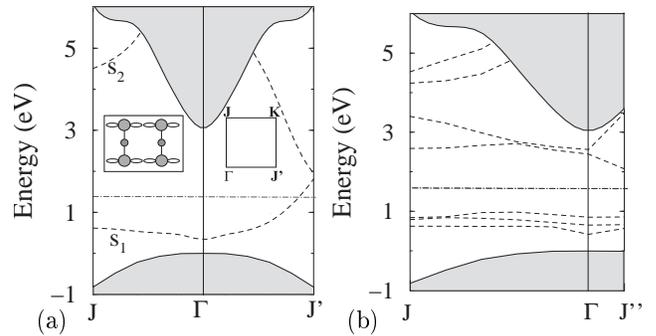,width=8.5cm,clip=f}
 \end{picture}
\caption{ Bandstructure calculated within the local density
  approximation for the relaxed Ga-terminated (a) (1$\times$1) and (b)
  (1$\times$4) surface of GaN.  The shaded region corresponds to the
  bulk projected bandstructure for zinc-blende GaN. The dashed lines
  are surface states.  In (a) the energetically lower surface state
  ($S_1$) is partially occupied with 3/2 electrons. The upper surface
  state ($S_2$) is empty. The dot-dashed line marks the position of
  the Fermi energy. In (b) the lower three surface states are doubly
  occupied; the upper surface states are empty.  $\overline{\Gamma
  {\rm J}^\prime}$ is parallel to the $[1\bar 10]$ direction.}
\label{fig:project_band}
\end{figure}

The atomic geometry for the unreconstructed Ga-terminated surface is
characterized by small vertical relaxations: The top layer (Ga)
relaxes 0.08\,\AA{} outward, the relaxation of the second layer (N) is
already negligible.  Surface relaxation can thus be excluded as a
possible origin for the unexpectedly large stability of this surface.

The calculated electronic bandstructure for this surface is shown in
Fig.~\ref{fig:project_band}(a).  For the unreconstructed surface our
calculations show two surface states ($S_1$, $S_2$). The energetically
lower surface state is a bonding state between neighboring surface Ga
atoms. The Ga-Ga bonds are {\em parallel} to the $[1\bar 10]$
direction explaining the large dispersion along the $\overline{\Gamma
{\rm J}^\prime}$ axis.  The lower surface state ($S_1$) is
partially occupied with 3/2 electrons rendering this surface
metallic. A comparison with the corresponding GaAs surface -- the
unreconstructed Ga-terminated (001) surface -- shows a qualitatively
similar bandstructure. The main difference is a significantly larger
dispersion (more than 1\,eV) of the surface bands on the GaN surface
reflecting the above mentioned formation of Ga-Ga bonds. The origin of
the large interaction lies in the sizable mismatch of the covalent
radii of Ga and N. Due to the small radius of the N atoms the Ga atoms
in GaN have approximately the same distance as in Ga
bulk~\cite{mrs94_s}. The Ga atoms at the surface can form metallic
bonds similar to those in bulk Ga even {\em without} any relaxation,
thus stabilizing the Ga-terminated surface.

However, a detailed analysis of the energies showed that the formation
of Ga-bonds on the surface is not sufficient to explain the
exceptionally low surface energy.  A second mechanism stabilizing
Ga-terminated surfaces with respect to N-terminated surfaces is the
very different cohesive energies of the bulk phases of Ga and N: The
cohesive energy of bulk Ga is 2.81\,eV/atom while that of the N$_2$
molecule is 5.0\,eV/atom. (The N--N bond in the N$_2$ molecule is one
of the strongest bonds found in nature.) In contrast the cohesive
energy of bulk As (2.96\,eV/atom) is only slightly larger than that of
bulk Ga.  Because of this asymmetry, more energy is required to
transfer N atoms from the N-reservoir to the surface than to transfer
Ga atoms to the surface.

The strong dispersion along the $\overline{ \Gamma{\rm J}^\prime}$
direction and the metallic character suggest that the Ga-terminated
$(1\times1)$ surface might be Peierls unstable along the [$1\bar 10$]
direction. Electron counting considerations indicate that the smallest
unit cell which allows an energy gap is a $(1\times4)$ structure.  In
fact, our calculations show that the $(1\times1)$ surface is unstable
against formation of a $(1\times4)$ reconstruction. The reconstructed
surface is semiinsulating with a bandgap of 1.2\,eV [see
Fig.~\ref{fig:project_band}(b)].  The atomic geometry is characterized
by {\em linear Ga tetramers} [see Fig.~\ref{fig:ball_and_stick}(a)].
The energy gain is 1.1\,eV per $(1\times4)$ cell compared to the
unreconstructed surface. The three bonds in the tetramer give rise to
three almost dispersionless surface states close to the valence band
which are each doubly occupied. The two dangling-bond orbitals [see
Fig.~\ref{fig:ball_and_stick}(a)] are unoccupied and give rise to the
two upper surface states [Fig.~2(b)].

A surface consisting of linear tetramers is unique and has neither
been observed experimentally nor studied theoretically for other III-V
(001) surfaces. Our calculations show indeed that on the corresponding
GaAs surface the tetramers are unstable and spontaneously form
dimers. The very different stability of tetramers on the two materials
originates again from the sizable mismatch of the covalent radii of Ga
and N. In order to form a tetramer on GaN breaking of back bonds is
not required and the Ga-N bond length remains almost unchanged
[$<0.1$\,\AA{}; see Fig.~3(a)]. On GaAs, however, forming a tetramer
is not possible without breaking back bonds. 

\setlength{\unitlength}{1mm}
\begin{figure}[tb]
 \begin{picture}(75,50)(0,-3)
\psfig{file=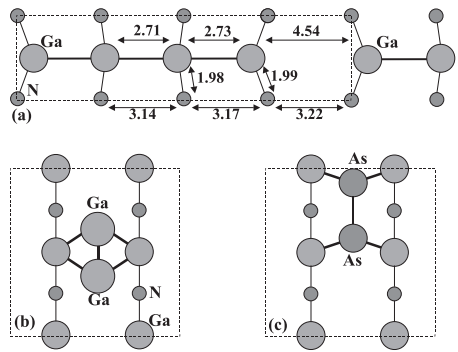,width=6cm,clip=f}
 \end{picture}
\caption{ Top view of the various surfaces: a) Ga-terminated
$(1\times4)$ (side view), b) $(2\times2)_{\rm Ga-dimer}$ and c)
$(2\times2)_{\rm As-dimer}$. The numbers give the
distance between atoms in \AA{}. As reference: The nearest and next
nearest neighbor distances in GaN are 1.95\AA{} and 3.17\AA{}
respectively.  }
\label{fig:ball_and_stick}
\end{figure}

In order to identify the experimentally observed reconstructions we
have studied a large number of possible configurations with different
surface stoichiometries and starting from very different initial
geometries and symmetries. These calculations will be described
elsewhere in more detail~\cite{to_be_published2}. Initially we focused
on structures which exhibit low surface energies on other III-V (001)
surfaces and which obey the electron counting rule. Examples are the
$c(2\times2)$ Ga and N vacancy structure and the $\beta2(4\times2)$
structure. However, all these surfaces are energetically less
favorable than the Ga-terminated $(1\times4)$ structure.

We then focused on the 
experimentally observed (2$\times$2) periodicity. STM measurements by
Wassermeier {\it et al.} indicate that the surface consists of {\it
one} dimer per $(2\times2)$ cell. During the initial stages of our
investigations we focused on clean GaN surfaces, but discovered that
all possible dimer structures are energetically higher in energy than
the tetramer structure. Starting with N dimers resulted always in the
formation of N$_2$ molecules which are bound by less than 50\,meV,
i.e. in a physisorbed state. Ga-dimers are energetically most stable
on a Ga-terminated surface [the geometry is shown in Fig. 3(b)];
however, the energy gain is too small to make them favorable [see
Fig. 1(b)]. We performed these calculations not only for the standard
dimer geometry but also for rotated and translated dimers. On the
basis of these results we conclude that there is no stable
$(2\times2)$ dimer structure on {\em clean} GaN (001) surfaces. We
emphasize, however, that the tetramer structure found here to be
stable is a strong candidate to explain the $(1\times4)$ RHEED pattern
observed by Feuillet {\it et al.}\cite{feuillet97}.

Not finding a stable $(2\times2)$ surface and in view of recent
experiments\cite{feuillet97} we studied the influence of arsenic on
the surface composition and structure.  We find that the energetically
most stable structure is a $(2\times2)$ structure with an arsenic
dimer on a Ga-terminated surface. Each As atom has two backbonds to
the underlying Ga atoms [see Fig.~3(c)].  The As-As dimer length is
2.46\,\AA.  Other structures with As in the second or third surface
layer, with mixed Ga-As bonds, and with higher coverages of As atoms
have been found to be energetically less stable. The surface energy of
the $(2\times2)$ As dimer structure is plotted in Fig.~1(b). Though
arsenic-poor conditions ($\mu_{\rm As}=\mu_{\rm As(bulk)}-\Delta
H_{\rm GaAs}$) have been assumed, the As-covered surface is
energetically clearly more favorable than the clean GaN (001)
surfaces, i.e. removing As from GaAs and putting it on GaN is an {\em
exothermic} reaction. The only assumption made here is that the system
is in thermodynamic equilibrium with GaAs. Considering the
comparatively large As vapor pressure of GaAs at typical GaN growth
temperatures this should be the case if growing on GaAs substrate or
in an As-contaminated growth chamber. We thus conclude that the
chemisorption of As atoms can significantly reduce the surface energy.

Our results further show that structures with As in the second or
deeper surface layer are energetically less favorable than structures
where As stays in the top surface layer. This result is consistent
with previous experimental\cite{weyers92} and theoretical
studies\cite{prb1_95} according to which As is virtually immiscible
in GaN. Both results, the reduction in surface energy and the low
miscibility suggest that As might be a good surfactant for the growth
of zinc-blende GaN. Generally, we expect also for other group V elements
a similar behavior. In particular, if As incorporation should turn out to 
be a problem using elements with an atomic radius {\em larger} than As 
(e.g. Sb, Bi) would be an interesting alternative.

Our results for clean and As covered surfaces allow some interesting
conclusions concerning the growth of GaN (001). First, both for clean
and As covered GaN the surface energy drops if going from N-rich
conditions toward Ga-rich conditions. The reason is that on both
surfaces the number of Ga atoms exceeds the number of N atoms: The
stable surfaces are always Ga-rich. This behavior is very different
from what has been found for the non-polar surfaces which are
stoichiometric except for extremely Ga-rich
conditions~\cite{northrup96}. Second, GaN (001) has a significantly
higher surface energy than the non-polar GaN surfaces implying that
the polar (001) surface is less stable. The difference in surface
energies {\em increases} and the stability of the (001) surface {\em
decreases} when going towards more N-rich conditions. This behavior
might explain why the growth morphology of cubic GaN severely degrades
when growing under N-rich conditions. It also casts severe doubts on
the old paradigm of pushing growth towards extreme N-rich conditions
in order to avoid a N deficiency. Instead, from a thermodynamic point
of view we expect best growth conditions under more Ga-rich conditions
where the surface energy is low. Finally it should be mentioned that
the large difference in atomic radii between Ga and N atoms and the 
extreme binding energy of
N$_2$, which drive the formation of Ga-rich surface structures,
is a general feature of the III-nitride semiconductors. For AlN, InN
and its alloys with GaN we expect therefore a similar behavior.

We acknowledge stimulating discussions with O. Brandt and M.
Wassermeier.  We gratefully acknowledge support from German Science
Foundation (JN), DARPA under agreement no.  MDA972-96-3-014 (CVdW) and
the BMBF and ``Fond der Chemischen Industrie'' (TZ).


\end{document}